\documentclass[journal=nalefd,manuscript=letter,layout=twocolumn]{achemso}
\setkeys{acs}{usetitle = true}
\usepackage[version=3]{mhchem} 
\let\oldmaketitle\maketitle
\let\maketitle\relax
\usepackage{multicol}
\usepackage{abstract}
\usepackage{subcaption}
\usepackage[labelformat=parens,labelsep=quad,skip=3pt]{caption}
\usepackage{graphicx}
\usepackage{float}
\usepackage[export]{adjustbox}
\usepackage{array}

\title{Design, Fabrication and Characterization of nanoplasmonic lattice for trapping of ultracold atoms}

\author{Sunil Kumar}%
\affiliation{Department of Physics, Indian Institute of Science Education and Research, Dr. Homi Bhabha Road,
Pashan, Pune 411 008
INDIA}%
\author{Manav Shah}
\affiliation{Tata Institute of Fundamental Research, Homi Bhabha Road, Mumbai 400005, India}
\author{Ajith P. Ravishankar}
\affiliation{Tata Institute of Fundamental Research, Homi Bhabha Road, Mumbai 400005, India}
\author{Chetan Vishwakarma} 
\affiliation{Department of Physics, Indian Institute of Science Education and Research, Dr. Homi Bhabha Road,
Pashan, Pune 411 008
INDIA}
\author{Arindam Dasgupta} 
\affiliation{Department of Physics, Indian Institute of Science Education and Research, Dr. Homi Bhabha Road,
Pashan, Pune 411 008
INDIA}
\author{Jay Mangaonkar} 
\affiliation{Department of Physics, Indian Institute of Science Education and Research, Dr. Homi Bhabha Road,
Pashan, Pune 411 008
INDIA}
\author{Venu Gopal Achanta} 
\affiliation{Tata Institute of Fundamental Research, Homi Bhabha Road, Mumbai 400005, India}
\author{ Umakant D. Rapol}
\email {umakant.rapol@iiserpune.ac.in}
\affiliation{Department of Physics, Indian Institute of Science Education and Research, Dr. Homi Bhabha Road,
Pashan, Pune 411 008
INDIA}
\altaffiliation {Center for energy sciences, Indian Institute of Science Education and Research, Dr. Homi Bhabha Road,
Pashan, Pune 411 008
INDIA}

\begin{document}
	\twocolumn[
	\oldmaketitle
	\begin{onecolabstract}
		
Ultracold atom-traps on a chip enhances the practical application of atom traps in quantum information processing, sensing, and metrology. Plasmon mediated near-field optical potentials are promising for trapping atoms. The combination of plasmonic nanostructures and ultracold atoms has the potential to create a two dimensional array of neutral atoms with lattice spacing smaller than that of lattices created from interfering light fields -- the optical lattices. We report the design, fabrication and characterization of a nano-scale array of near-field optical traps for neutral atoms using plasmonic nanostructures. The building block of the array is a metallic nano-disc fabricated on the surface of an ITO-coated glass substrate. We numerically simulate the electromagnetic field-distribution using Finite Difference Time Domain method around the nanodisc, and calculate the intensity, optical potential and the dipole force for $^{87}$Rb atoms. The optical near-field generated from the fabricated nanostructures is experimentally characterized by using Near-field Scanning Optical Microscopy. We find that the optical potential and dipole force has all the desired characteristics to trap cold atoms when a blue-detuned light-field is used to excite the nanostructures. This trap can be used for effective trapping and manipulation of isolated atoms and also for creating a lattice of neutral atoms having sub-optical wavelength lattice spacing. Near-field measurements are affected by the influence of tip on the sub-wavelength structure. We present a deconvolution method to extract the actual near-field profile from the measured data. 
	\end{onecolabstract}
	]

	 The study of atom-light interaction at the subwavelength scale using the combination of nanoplasmonics with ultracold atoms \cite{chang2006quantum} is an emerging field of research. Nanoplasmonic systems offer the possibility of generating trapping potentials for ultracold atoms in the optical near fields \cite{chang2009trapping,Stehle2011,Stehle2013,esteve2013cold}. This is based on the interaction of ultracold atoms with plasmonic excitations of  metal nanostructures. This kind of hybrid system can generate nanotraps to give precise control over the atomic motion and achieve strong coupling between single atoms and plasmonic nanostructures. The high degree of freedom over controlling the coupling and interaction enables this hybrid system to be a strong candidate for Hamiltonians of a) strongly correlated systems from condensed matter physics, b) cavity QED \cite{miller2005trapped} and for c) Quantum Information Processing (QIP). A major limitation faced by optical lattice traps for ultracold atoms is that the nearest neighbour tunneling rate is fundamentally limited (to $\sim$10s of kHz) by the wavelength of laser light used. This limitation can be overcome by the use of a lattice of plasmonic nanostructures, which exhibit subwavelength lattice spacing and strong near-field enhancement \cite{chang2009trapping,Stehle2011,Stehle2013,esteve2013cold,maier2007plasmonics}. The enhanced electromagnetic fields confined to sub-optical wavelength regime allows the creation of optical dipole force traps to confine atoms to regions below the optical diffraction limit. The integration of plasmonic excitations and ultracold atoms can boost up the tunneling rate by orders of magnitude as compared to a conventional optical lattice \cite{JuliaDiaz2013}.

	In this article we report the design, fabrication and optical characterization of plasmonic nanodiscs for trapping of ultracold sample of bosonic $^{87}$Rb atoms and analyze the feasibility of generating 2D array of optical potential for the simulation of condensed matter Hamiltonians with enhanced tunneling rate. In addition, these structures may also show promise in designing efficient nano-plasmonic structures for solar light harvesting. We start with the details on numerical simulations performed using Finite Difference Time Domain (FDTD) method to study the electric field profile around the nanodiscs, and optimise the material as well as geometrical parameters in order to achieve maximum enhancement and optimal dipole trapping potential. We discuss the fabrication of the optimised plasmonic nanostructure. Finally, we show results from Near-field Scanning Optical Microscopy on the fabricated samples in order to establish the near-field features showing the trap potential above the nanodisc.\par

\begin{figure}[h!]
  	\begin{subfigure}{8cm} 		
		\includegraphics[width=.95\textwidth]{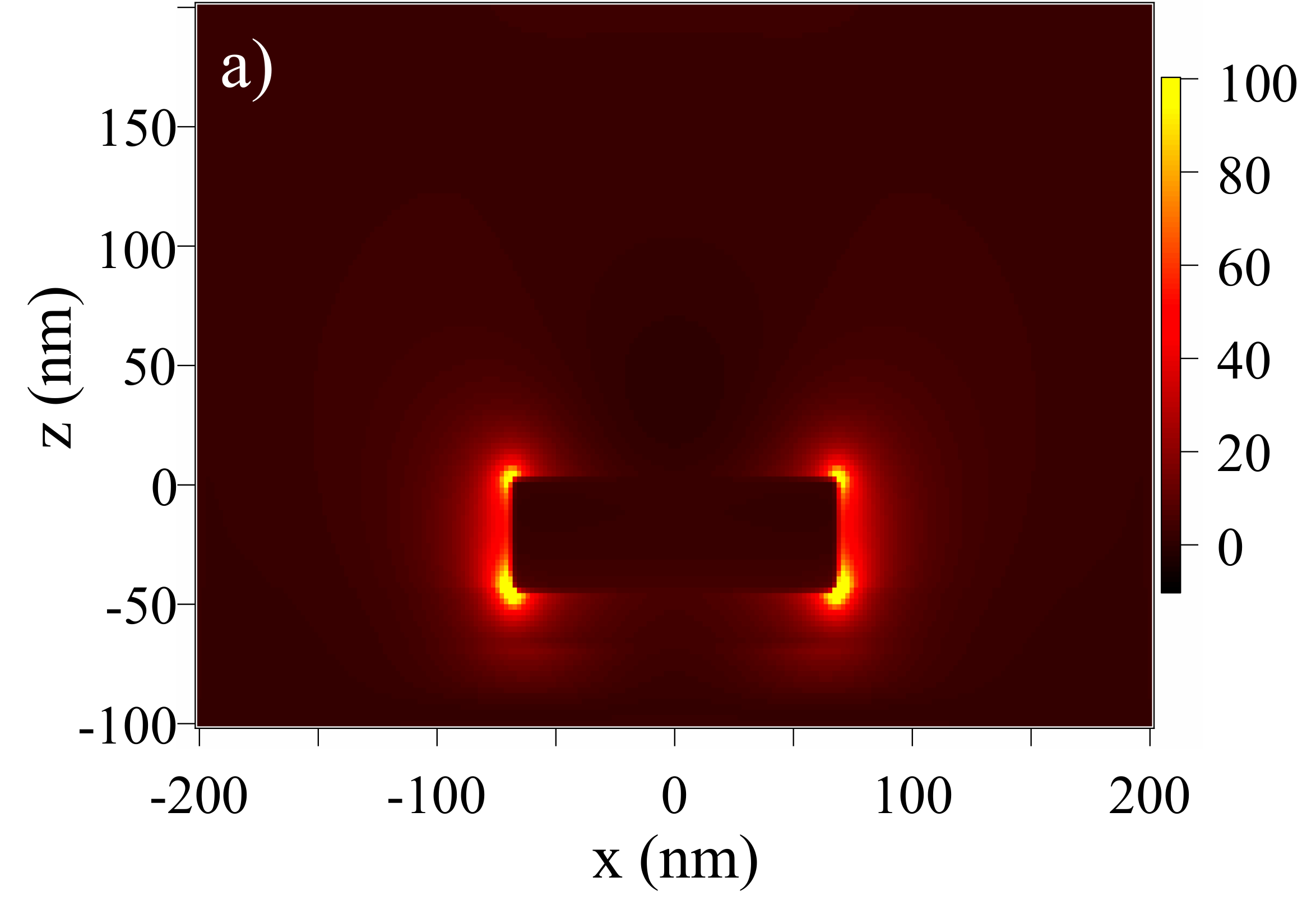}
  	\end{subfigure}
  	
	\begin{subfigure}{8cm}
   		\includegraphics[width=.8\textwidth]{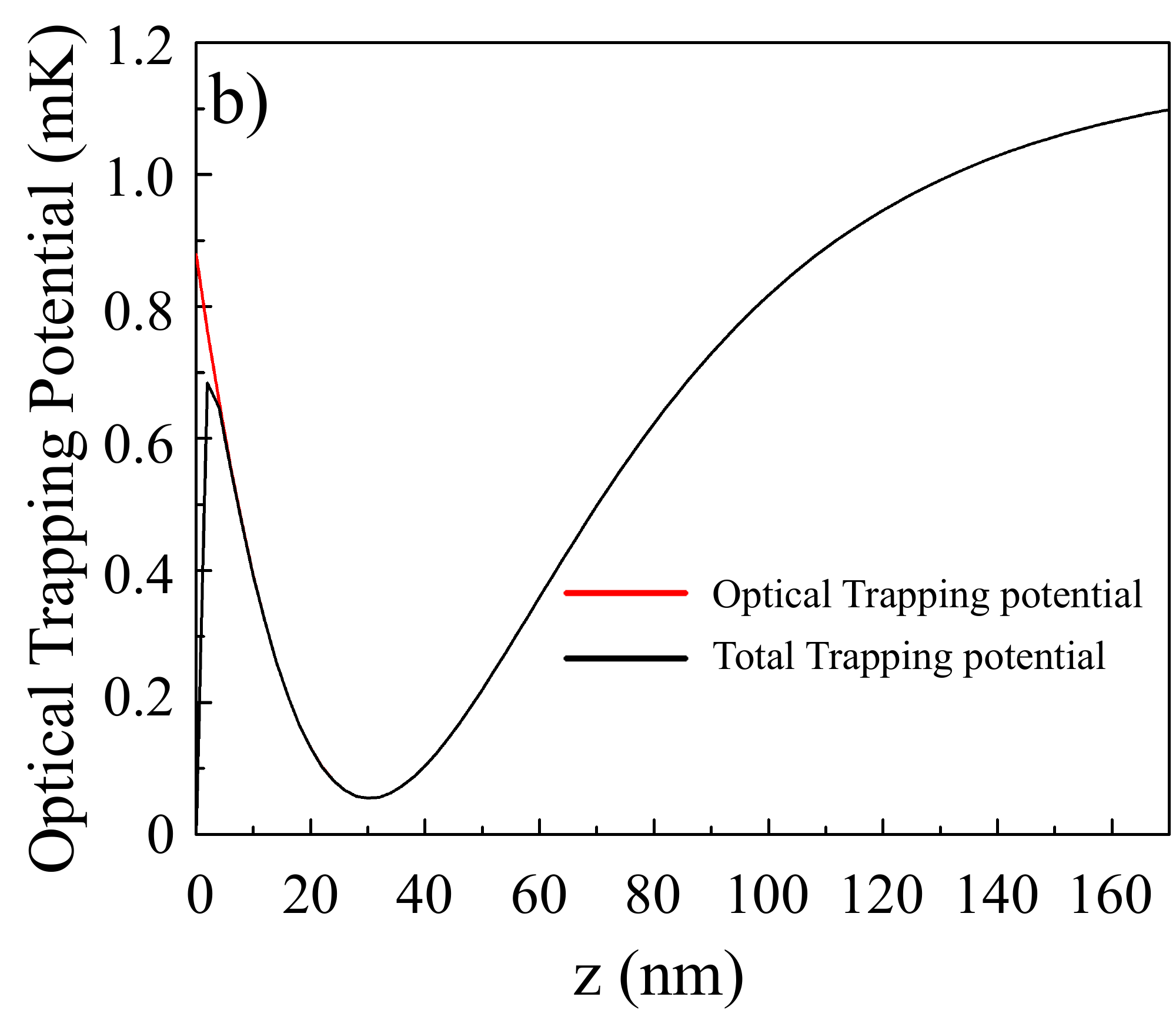}
 	 \end{subfigure}
	\caption{\textbf{Simulated electric-field intensity around a nanodisc:} Excitation by circularly polarised light leads to Localised Surface Plasmon - enhanced field confinement along the periphery of the nanodisc. a) shows a 2D plot of the electric field intensity with a cross-section across the disc in the xz-plane. The darkest region representing the minima in the optical field forms the trapping potential. b) Shows a the section of the optical potential in the z-direction above the center of the disc in a). The black curve shows the sum of the Van der Waals potential and the optical potential. The potential is dominated by the optical potential.}
	\label{fig1}
\end{figure}

	Scattering of light from metallic nanodiscs gives rise to Localised Surface Plasmon modes that lead to field confinement around its periphery\cite{imura2014plasmon}. Such an intensity profile can be exploited to trap ultracold neutral atoms, molecules and nanoparticles \cite{garciasegundo2007, kang2014plasmonic}. For linearly polarised light, the plasmonic near field exhibits a dipolar intensity profile. Instead, exciting a nanodisc using circularly polarised light leads to confinement of field along its edge\cite{wang2011trapping} (fig.\ref{fig1}). The electric field is found to be greatly enhanced around its periphery and is out of phase with the incident electric field. A minimum of field intensity (z$_{min}$) is found at a distance of a few 10's of nanometers along z axis from the disc surface and can be tuned by changing the laser parameters and the dimensions of the nanodisc. For a simple two level atom, the electric field profile (in this case a minimum in the center) provides a trapping potential when the laser frequency is blue detuned with respect to the transition frequency. Such an intensity profile can act as an optical dipole trap for neutral atoms.

\begin{figure}[ht!]
	\includegraphics[width=8cm]{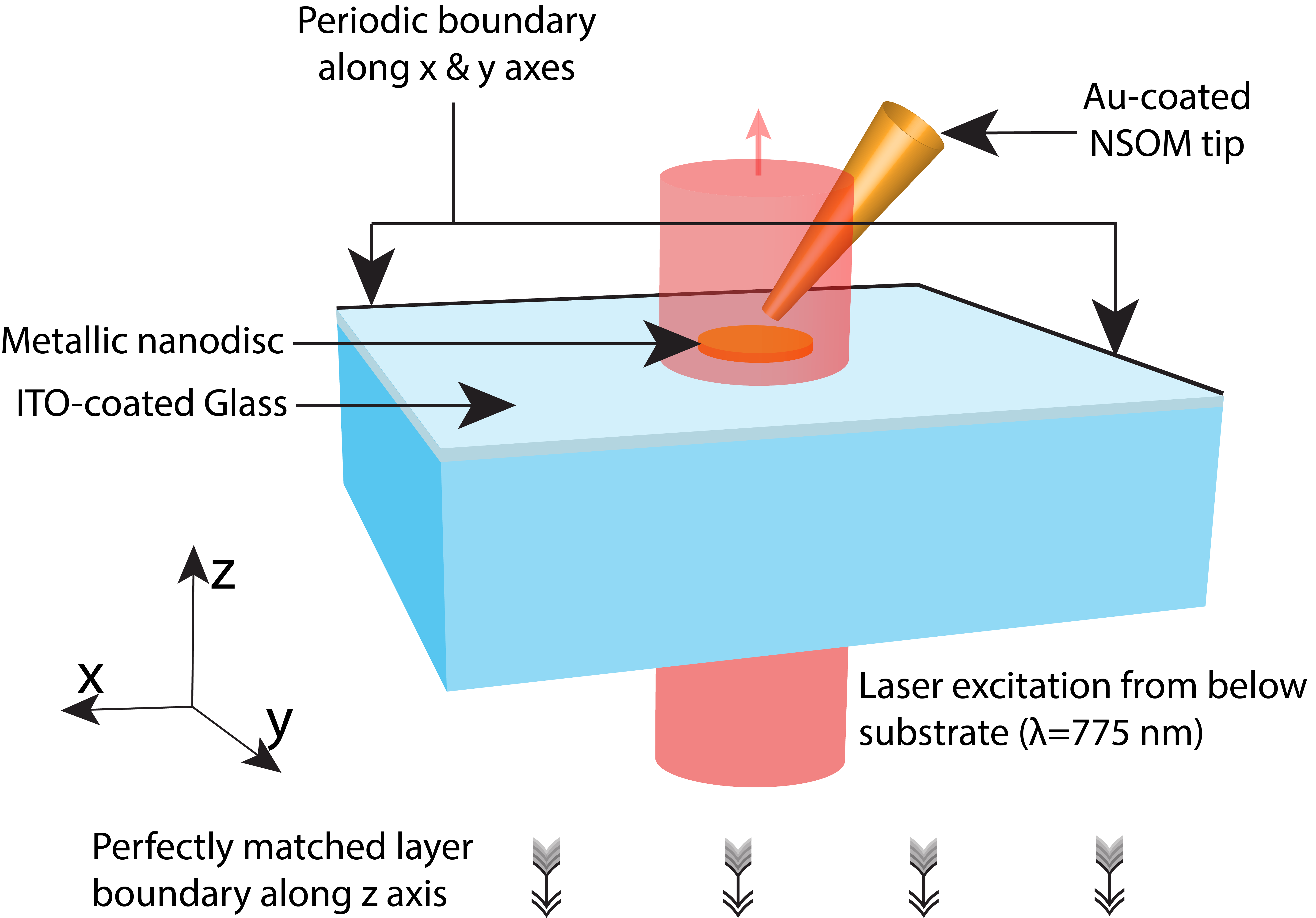}
	\caption{ \textbf{Simulation setup:} The schematic shows the simulation setup for the system. The nanodisc is excited by a gaussian plane wave light source of wavelength 775 nm from below the glass substrate.}
	\label{fig2}
\end{figure}

	 The simulation setup consists of an Indium Tin Oxide (ITO) - coated glass substrate, with 20 nm ITO coating, dressed with isolated nanodiscs (fig. \ref{fig2}). In order to find the optimum potential profile, we carried out numerical simulations by FDTD method on an isolated metallic nanodisc with perfectly matched layer (PML) boundary conditions as well as an array of nanodiscs with a period of 5 $\mu{m}$ (above the Rayleigh limit) with periodic boundary conditions (PBC). The grid size is 2 nm in the simulations. Gold(Au) and Silver(Ag) were used as the materials for nanodisc due to their favorable plasmonic response.  The wavelength dependent dielectric constants for gold and silver were taken from experimental values provided by Johnson and Christy \cite{johnson1972optical}. The diameter (D) of the nanodiscs is varied from 100 nm - 250 nm and height (H) from 10 nm - 60 nm. Values below and above the selected range were avoided due to fabrication limitations and unfavorable intensity profile, respectively. A Total-Field Scattered-Field sources (TFSF) source for studying a single disc and a  plane wave source for periodic array of discs were used for excitation from below the substrate. The simulated electric field profile is recorded by placing 2D-field monitors on top of the nanodisc and across the nanodisc in the $y=0$ plane. For simulations without an NSOM tip, PBC is used along x and y axes to reduce computational effort, while the presence of tip requires PML boundary conditions. PML boundary is applied to z axis in all simulations. We found that using smaller D and H resulted in higher field enhancement, but shallow potential depth above the nanodisc, while higher D and H values resulted in z$_{min}$ being farther away from the nanodisc.\par

	Using the electric field extracted from fig.\ref{fig1}, the trapping potential in terms of temperature (T) can be calculated by\cite{GRM89}
\begin{equation}
	U_{op} = \frac{3\pi c^{2}}{2w_{0}^{3}} \frac{\Gamma}{\Delta}I(\mathbf{r})
\end{equation}
\begin{equation}
	T = \frac{U_{op}}{k_b}
\end{equation}
where $ \omega_0 = 2\pi c(780 nm)^{-1}$ is  the principle transition frequency, $\Gamma$ is the decay rate for $^{87}$Rb, $\Delta=\mid\omega_0-\omega\mid$ is the frequency detuning and $I(\mathbf{r})$ is the electric field intensity, which is taken to be 5$\times$10$^{7}$ W/m$^{2}$. 

\begin{figure}[H]
\begin{subfigure}{8cm} 
	\includegraphics[width=0.9\textwidth]{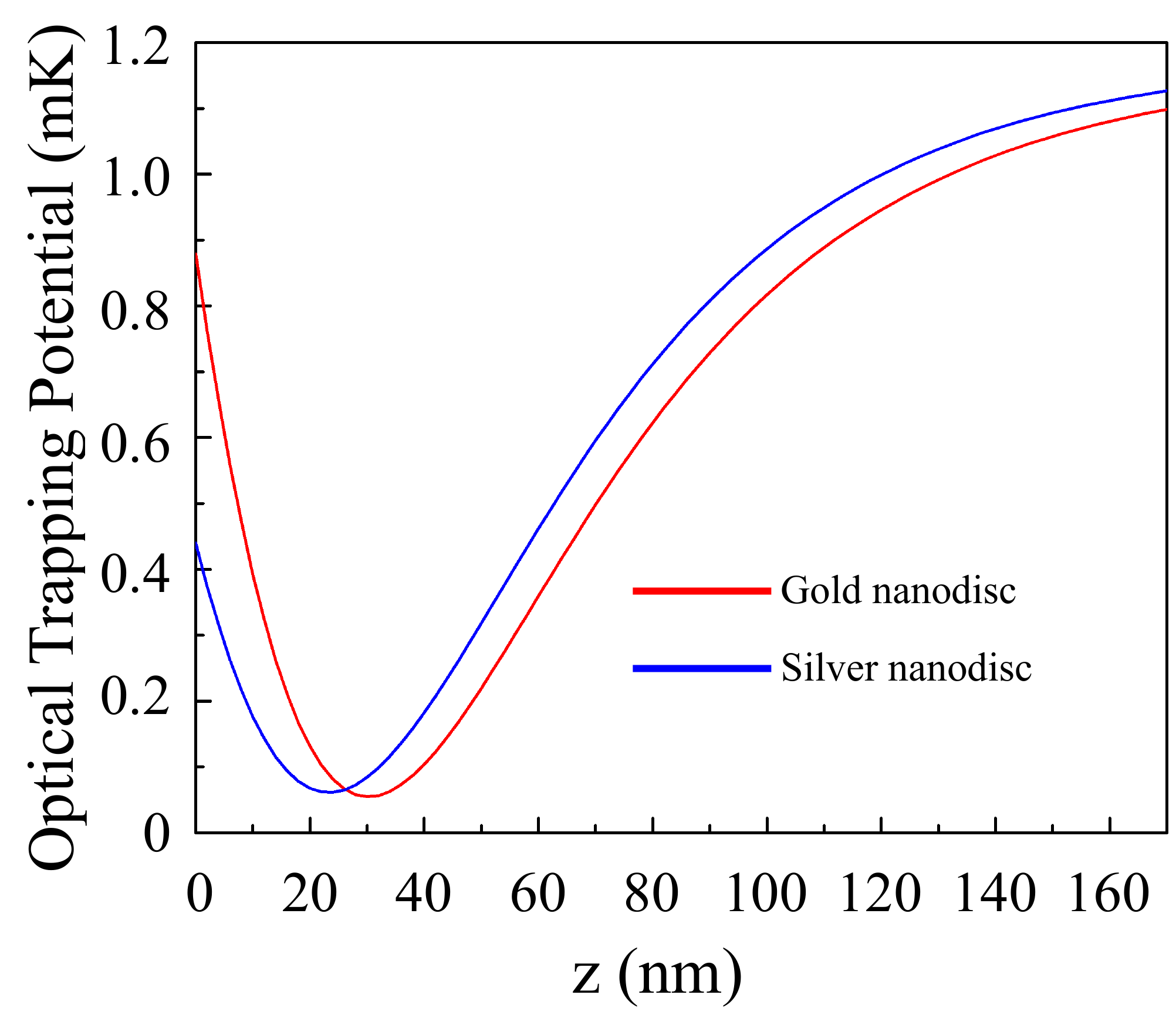}
\end{subfigure}
\caption{ \textbf{Trapping potential:} Simulated optical potential in z-direction for single Rb atom using a silver(blue) or gold(red) nanodisc of diameter 150 nm and 40 nm height. A potential minima is formed at a distance of ~30 nm from the surface of the nanodisc for gold and ~24 nm for silver nanodisc, which can act as a dipole trap.}
\label{fig3}
\end{figure}

Fig. \ref{fig3} shows the trapping potential in terms of temperature along the z-direction ($\mathbf{r}=z\hat{\mathbf{z}}$), taking the top of the nanodisc's surface as $z=0$. The resultant optical trap distance from surface of the nanodisc, trap frequency and depth are given in table \ref{table1}.

\begin{table}[h]
\centering
\footnotesize
\begin{tabular}{|m{1cm}|m{1.5cm}|m{1.5cm}|m{2cm}|} 
 \hline
Disc & Trap Distance (nm) & Trap Frequency (MHz) & Trap Depth (mK) \\ 
 \hline
 Silver & 24 & 18 & 0.49 \\ 
 Gold & 30 & 18 & 0.88 \\
 \hline
\end{tabular}
\caption{\textbf{Calculated parameters of optical potential trap for silver and gold nanodiscs from simulated field intensity.}}
\label{table1}
\end{table}

\begin{figure*}[h]
		\begin{subfigure}{8cm}
		\includegraphics[width=7cm]{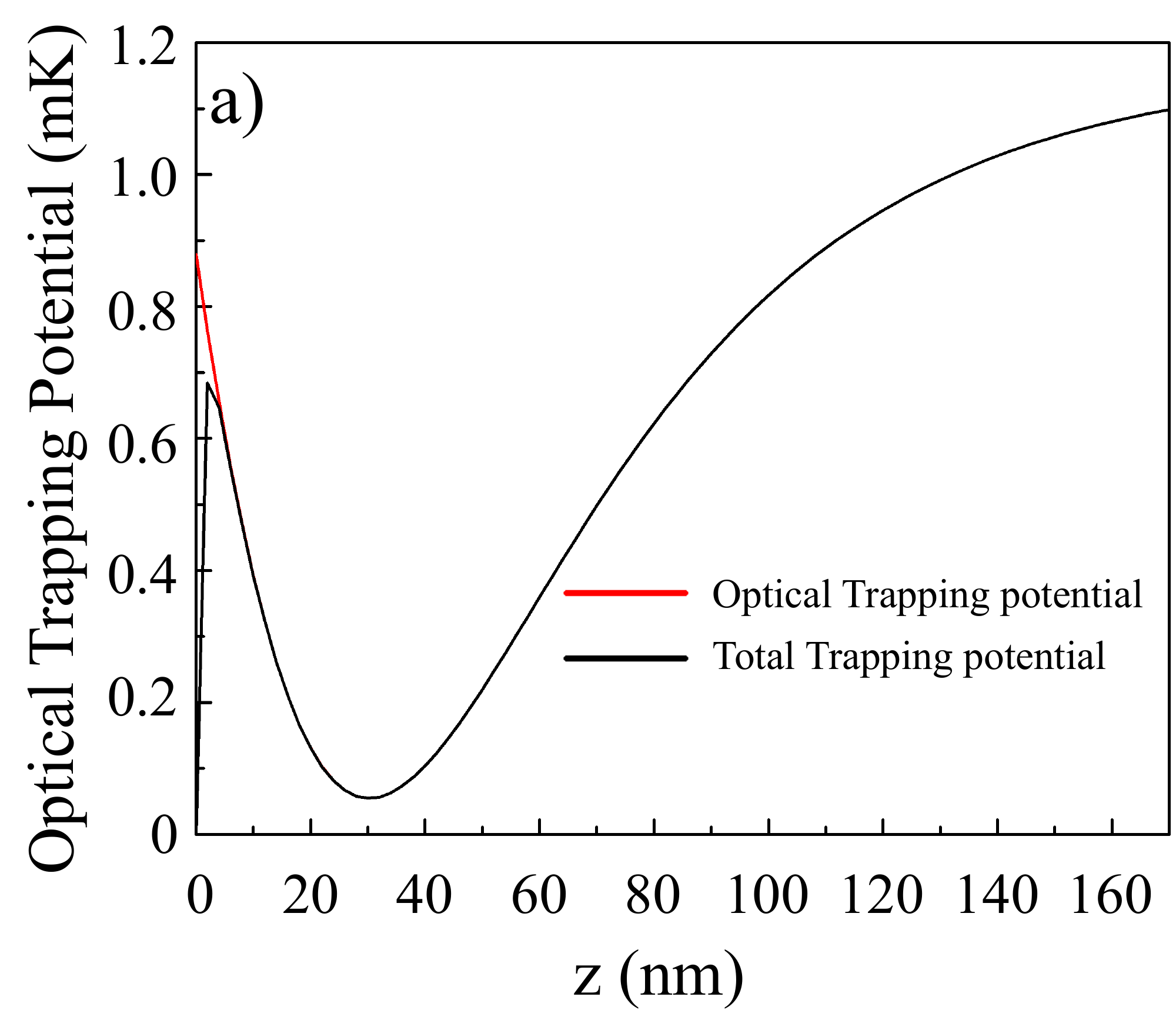}
		\caption{\label{fig4a}}
		\end{subfigure}
		\begin{subfigure}{8cm}
		\includegraphics[width=7cm]{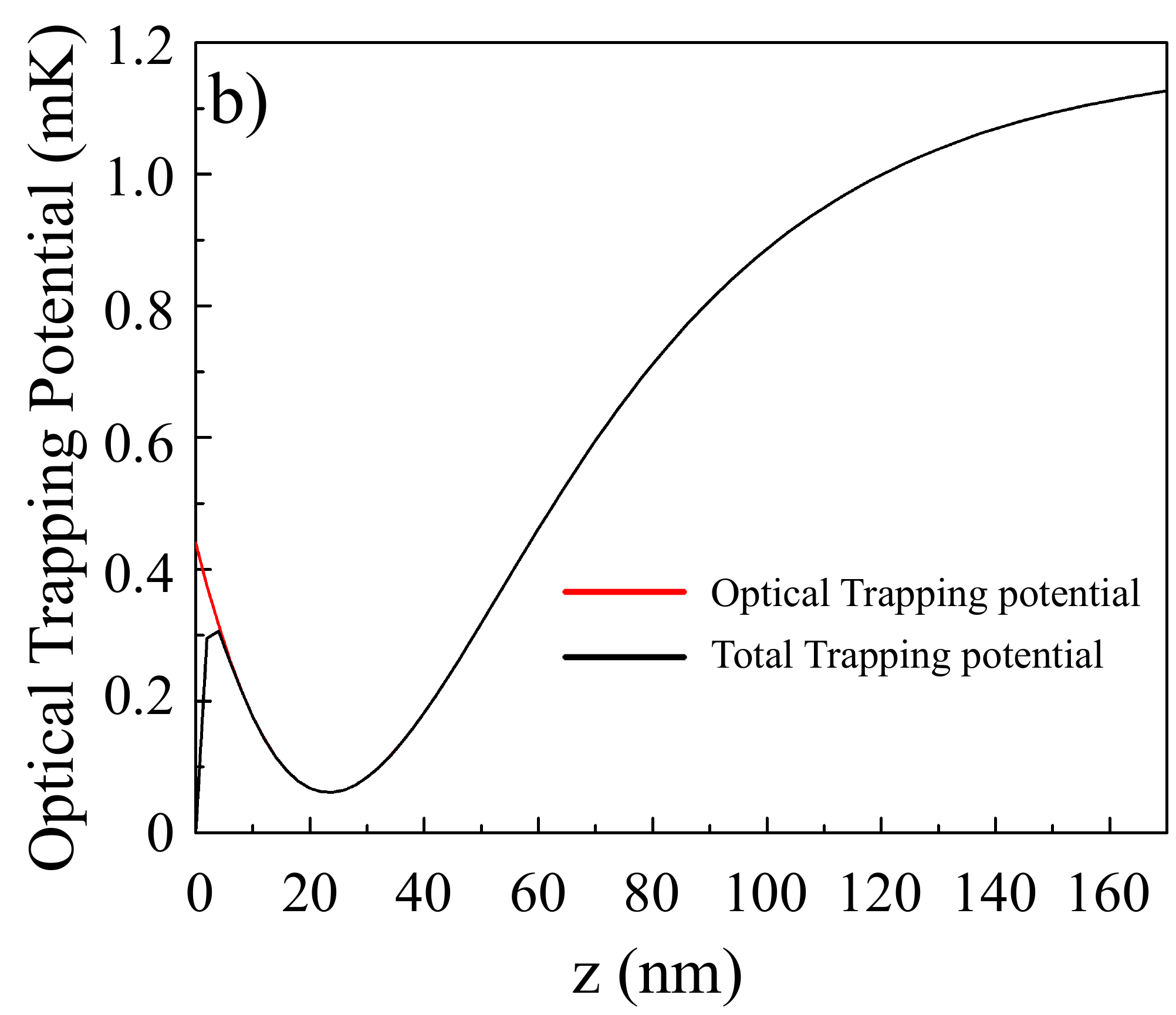}
		\caption{\label{fig4b}}
		\end{subfigure}
		\caption{ \textbf{Trapping potential:} Simulated \textbf{total} potential in z-direction for single Rb atom for a gold nanodisc (a) and silver nanodisc (b) of diameter 150 nm and height of 40 nm, modified due to Casimir-Polder (CP) potential. Red curve shows the optical dipole potential while the black curve shows the total potential.}
		\label{fig4}
\end{figure*}
	
	The distance between minima of the trap and the surface of the nano disc is  d $\simeq$ 20 nm which gives rise to trapping of atoms leading to atom - surface interactions due to attractive Van der Waals (VdW) force \cite{genet2003electromagnetic,JuliaDiaz2013,chang2009trapping}. Hence, to quantify this effect, the Casimir-Polder (CP) interaction with the surface is often used as an approximation and given by \cite{mihaljevic2014geometric, perreault2008modifying, daly2014nanostructured}:

\begin{equation}
	U_{vdW} =-\frac{C_3^{(0)}}{|z|^3}, \ \                     |z| << \lambda
\end{equation} 
	 In such a scenario, the total potential experienced by the atoms is a combination of optical potential and CP potential,
\begin{equation}
	U_{Total}= U_{opt}-\frac{C_3^{(0)}}{|z|^3}, \                  
	C_3^{(0)}=  \dfrac{e^2 \hbar}{32\pi\epsilon_0m_{Rb}\omega_0}
\end{equation}
which is valid for perfectly conducting metal surfaces (here, metallic nanodiscs). Here, $e$ denotes electronic charge, $\hbar$ is the reduced Planck's constant, $\epsilon_0$ is the permittivity of free space, and $m_{Rb}$ is the mass of a $^{87}$Rb atom. The optical potential must be strong enough to support a local potential minimum for trapping a cloud of atoms. The total potential is shown in fig.\ref{fig4}. Throughout the simulations, the effect of heating due to polarization noise of the laser beam has been neglected.\par
	
\begin{figure}[h]
		\includegraphics[width=5.5cm]{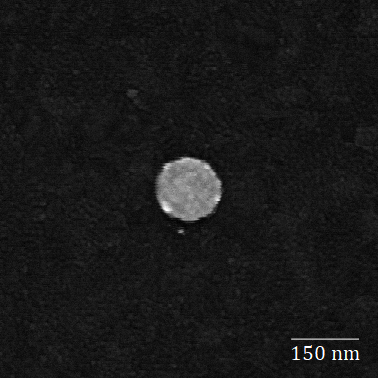}
		\caption{\textbf{Scanning Electron Micrograph of a fabricated gold nanodisc of diameter 150 nm.}}
		\label{fig5}
	\end{figure}

	After optimizing the trap parameters we fabricated the nanodiscs for optical characterization. The nanostructures were fabricated with very high repeatability and tight fabrication tolerance by `Top-Down' approach using Electron beam (E-beam) Lithography \cite{stewart2008nanostructured,gates2005new,hanarp2003control}. Glass slide of thickness 1.1 mm, with an ITO coating of 15-30 nm, was used as the substrate for sample fabrication. The use of ITO coated glass substrate ensures a minimum charging effect during the E-beam patterning. We used Polymethyl Methacrylate (PMMA) with a molecular weight of 996K, or AR-P 6200 (Allresist) as positive E-beam resists for lithography. The resist is spin-coated on a clean substrate, followed by baking to boil off the solvent. The nanodisc pattern is then directly written onto the resist by E-beam lithography. The exposed resist is dissolved in a suitable developer to produce required pattern on the resist. Au or Ag is then deposited on top of the resist by sputtering and thermal vapour deposition, respectively. Finally, the resist is removed via a lift-off process, leaving behind a pattern of isolated nanodiscs on the ITO-coated glass. Fig. \ref{fig5} shows a Scanning Electron Micrograph (SEM) of the fabricated Au nanodisc sample.

\begin{figure}
	\begin{subfigure}{8cm} 
	\includegraphics[width=0.95\textwidth]{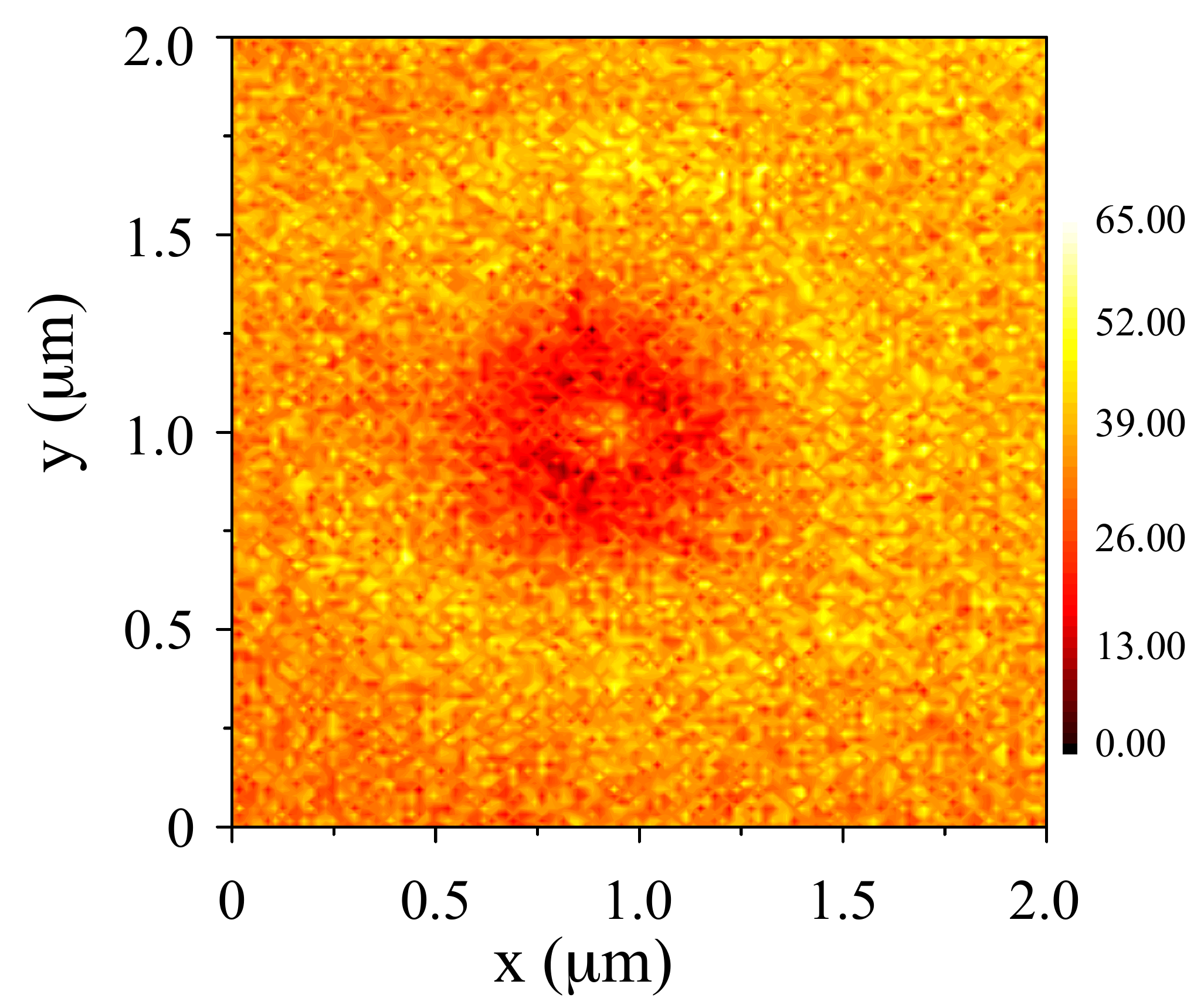}
\end{subfigure}
		\caption{\textbf{NSOM measurement of silver nanodiscs:} Figures show the experimentally measured near field with the coated fiber tip having an aperture of 150 nm at the apex with a scan window 2 $\mu $m$\times$2 $\mu $m. Due to resolution limitations of the NSOM, the central ring appears like a blurry central bright spot inside the red ring. Intensity is given in counts (kHz).}
		\label{fig6}
	\end{figure} 

	The optical trapping scheme using nanodiscs relies on their plasmonic response, which evanescently decays away from the surface of the sample\cite{novotny2012principles, maier2007plasmonics}. While far-field techniques are limited by diffraction limit, with near-field scanning optical microscopy (NSOM) it is possible to measure the optical properties of the plasmonic nanostructures with a spatial resolution in the range of few nanometres limited only by the tip dimensions \cite{novotny2012principles}.  We used Nanonics Multiview 2000 AFM/NSOM system with a bent,  Au-coated optical fiber probe with sub-wavelength aperture.  We used circularly polarized 775 nm wavelength from a supercontinuum laser (NKT Photonics) as source to excite the sample from below the substrate. The near-field is measured in transmission mode, with the tip raster-scanning over the nanodisc in `tapping' mode. The intensity collected via the aperture of the tip is detected by an avalanche photodiode (APD) in a pulse counting mode. The captured intensity at every spatial point is used to get an image. We used fiber tips with 150 nm and 200 nm tips. The obtained result for 150 nm disc diameter is shown in the Fig.\ref{fig6}.
		
\begin{figure}[t]
\begin{subfigure}{8cm} 
		\includegraphics[width=0.95 \textwidth,angle=0]{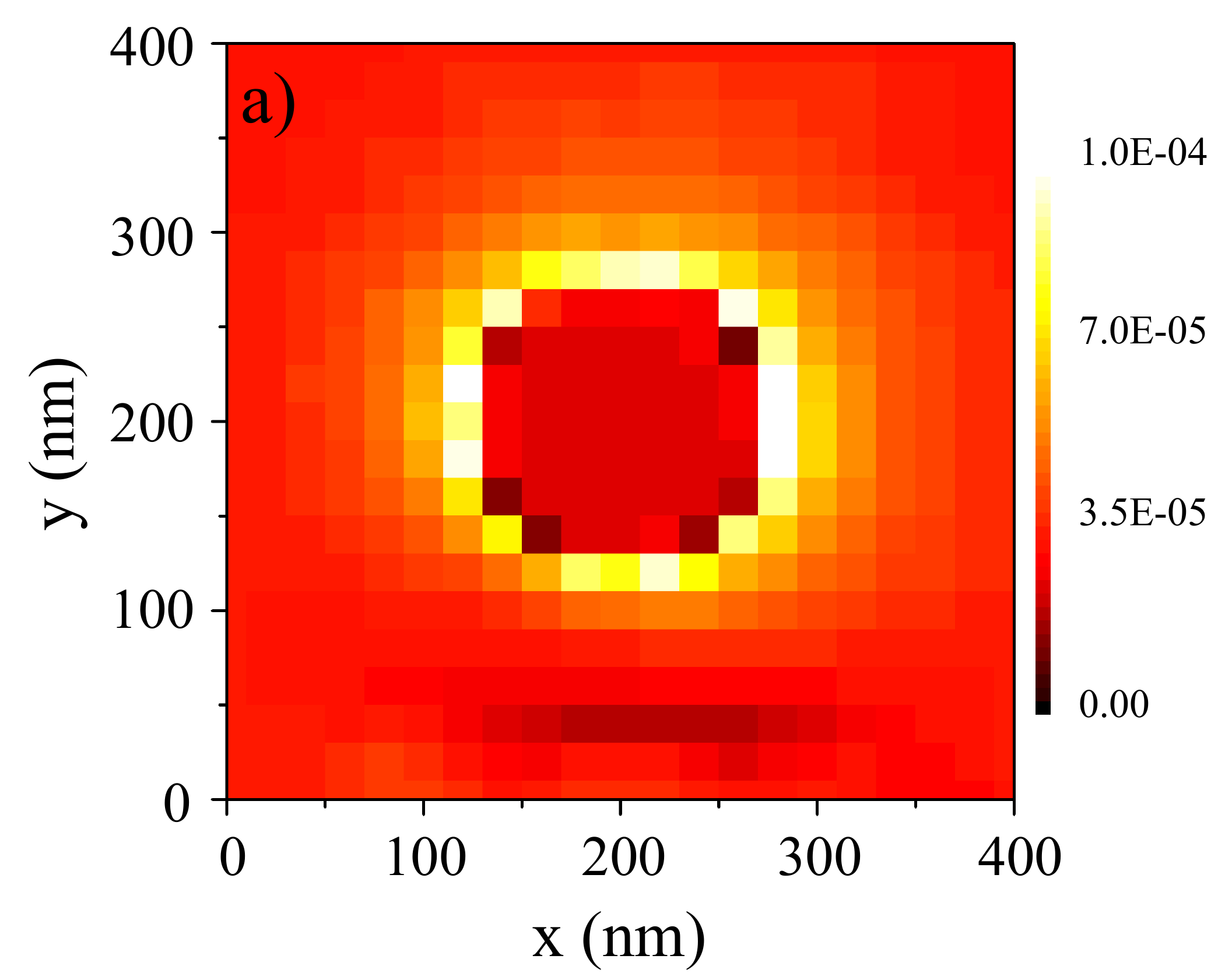}
		\includegraphics[width=0.95 \textwidth,angle=0]{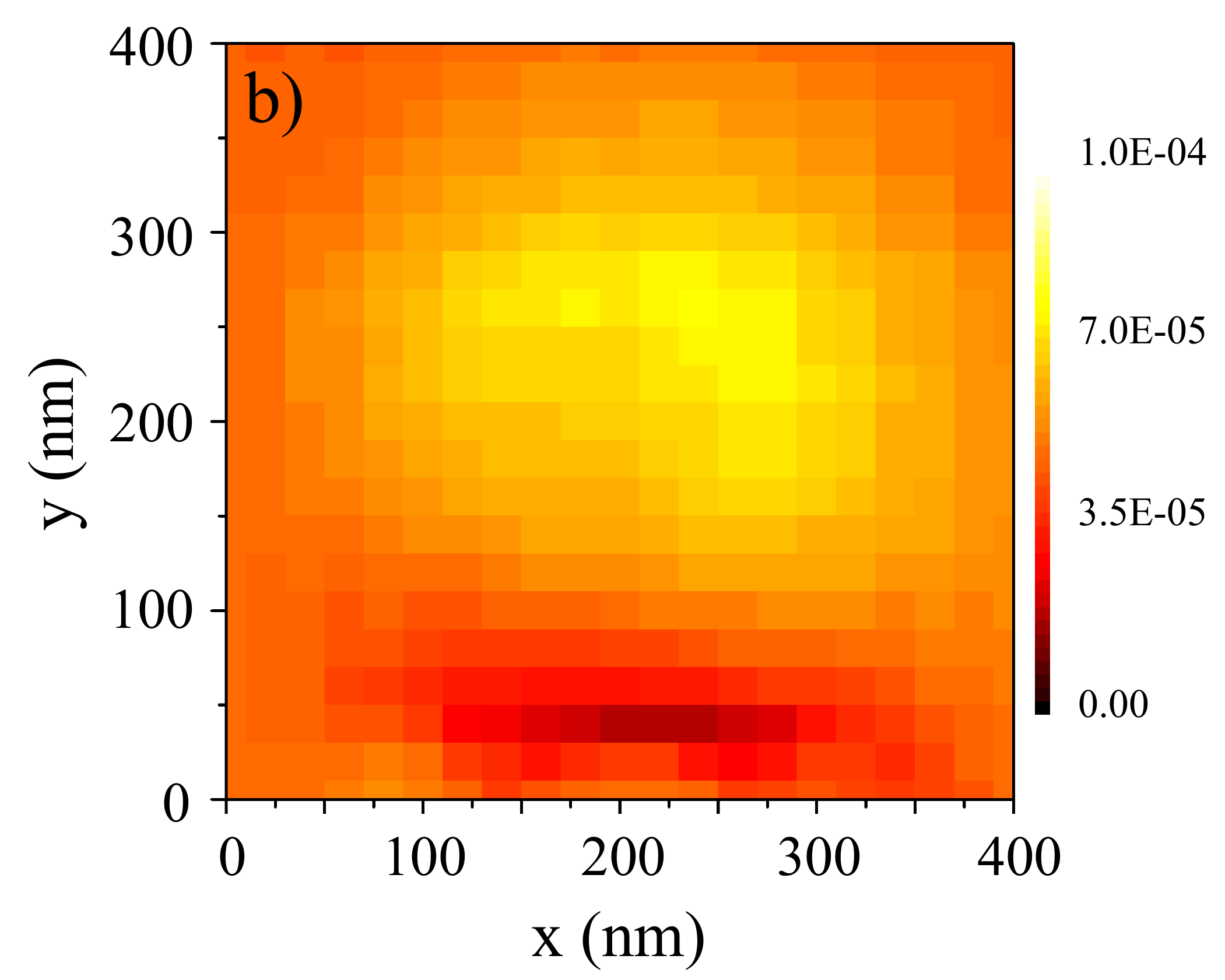}
\end{subfigure}
		\caption{\textbf{Effect of gold coated fiber tip on the NSOM scan:} Figure showing \textbf{simulated} NSOM scan with gold coated fiber tip over the nanodisc. The image is constructed by taking the E - field intensity at each tip position.The intensity captured at two different heights- a) at the surface of nanodisc b) at the centre of the NSOM tip.}
	\label{fig8}
	\end{figure}

	While NSOM technology enables optical characterisation beyond the diffraction limit, it suffers from a major limitation. The introduction of a sub-wavelength probe into the near-field of the sample distorts the natural near-field of the sub-wavelength structure. The recorded images with NSOM are often different from the actual near-field intensity profiles due to the strong interaction between sample structure and the probe\cite{novotny2012principles}. In the present case, the NSOM tip coated with Au  perturbs the near-field of the sample. Hence, the observed images (see Fig.\ref{fig6}) differ from the expected near field response with the ideal conditions as shown in Fig.\ref{fig8}a. Additionally, the resolution of the NSOM scanning is limited by the speed (defined by acquisition time per point) of the scan, oscillation of the tip, and the tip size. Though the first two factors can be easily controlled by changing the parameters of the scanning, the tip size limits the resolution. To extract the actual near-field profile, one needs to deconvolute the effect of tip from the measured near-field. In the following we present a deconvolution method.

\begin{figure*}[hbt!]
		\begin{subfigure}{6.0cm}
		\includegraphics[width=0.9 \textwidth]{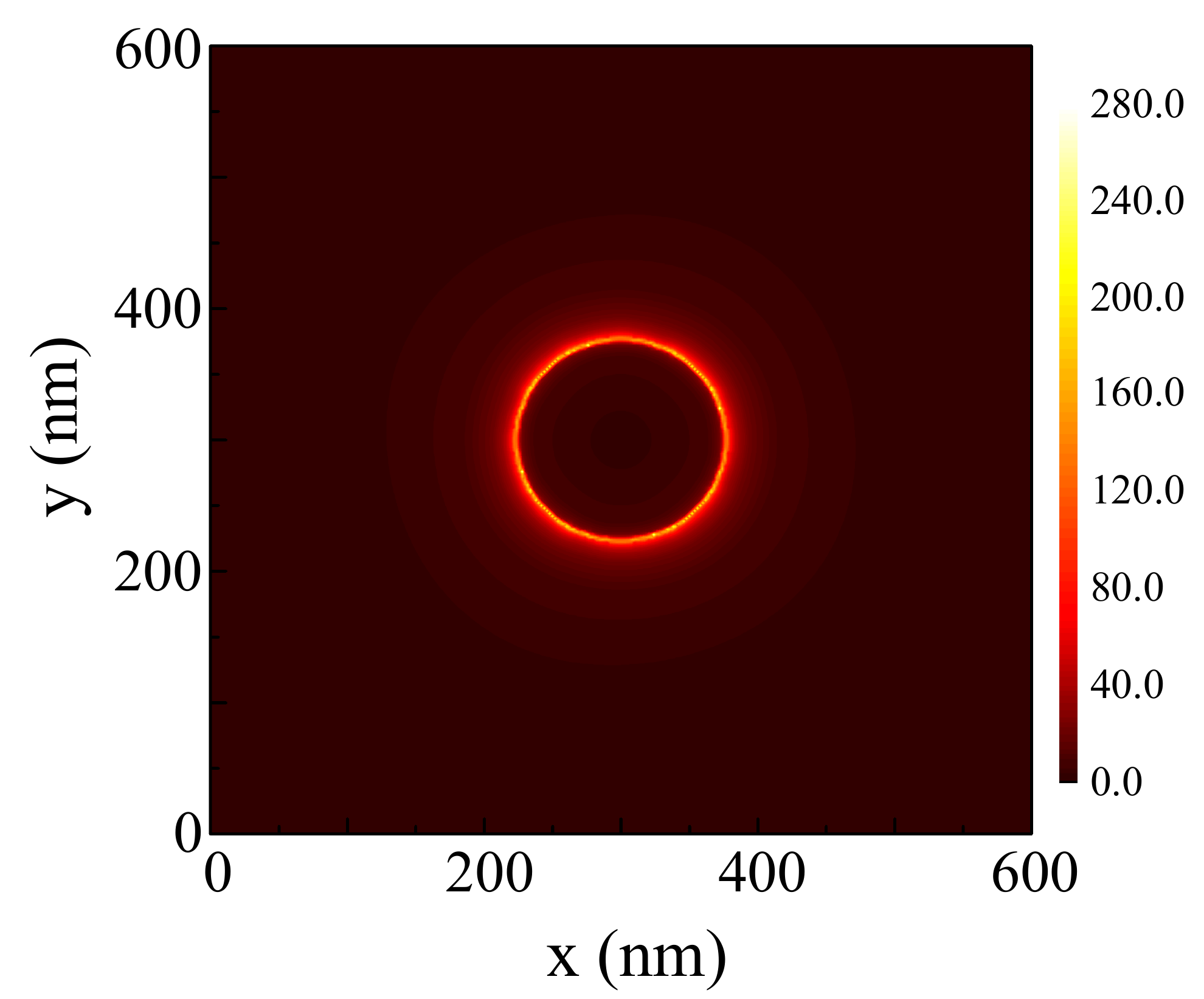}
		\caption{\label{fig9a}}
		\end{subfigure}
\begin{subfigure}{6.5cm}
		\includegraphics[width=0.8 \textwidth]{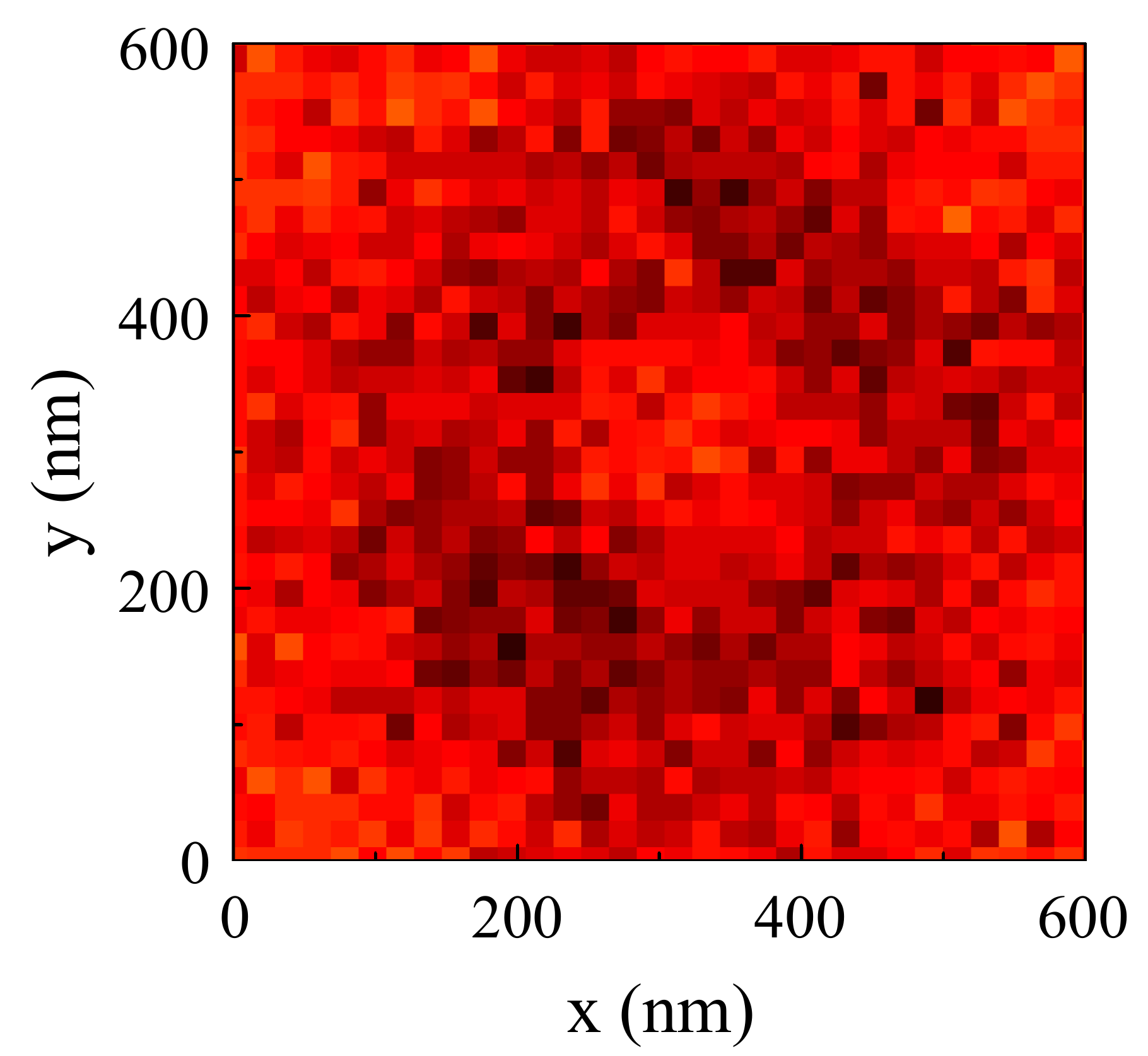}
		\caption{\label{fig9b}}
		\end{subfigure}
\begin{subfigure}{5.0cm}
		\includegraphics[width=0.95 \textwidth]{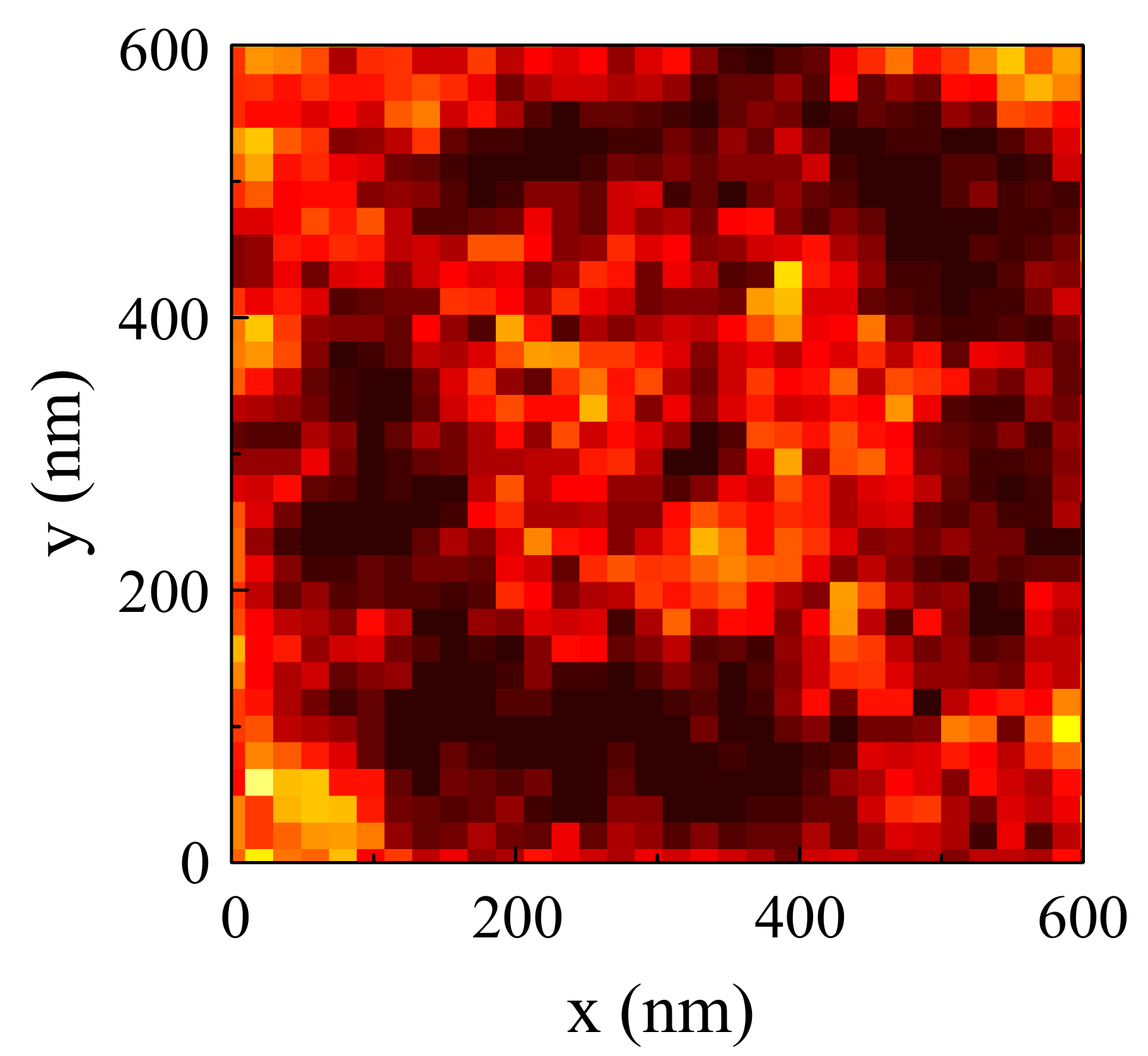}
		\caption{\label{fig9c}}
		\end{subfigure}
\begin{subfigure}{5.0cm}
		\includegraphics[width=0.95 \textwidth]{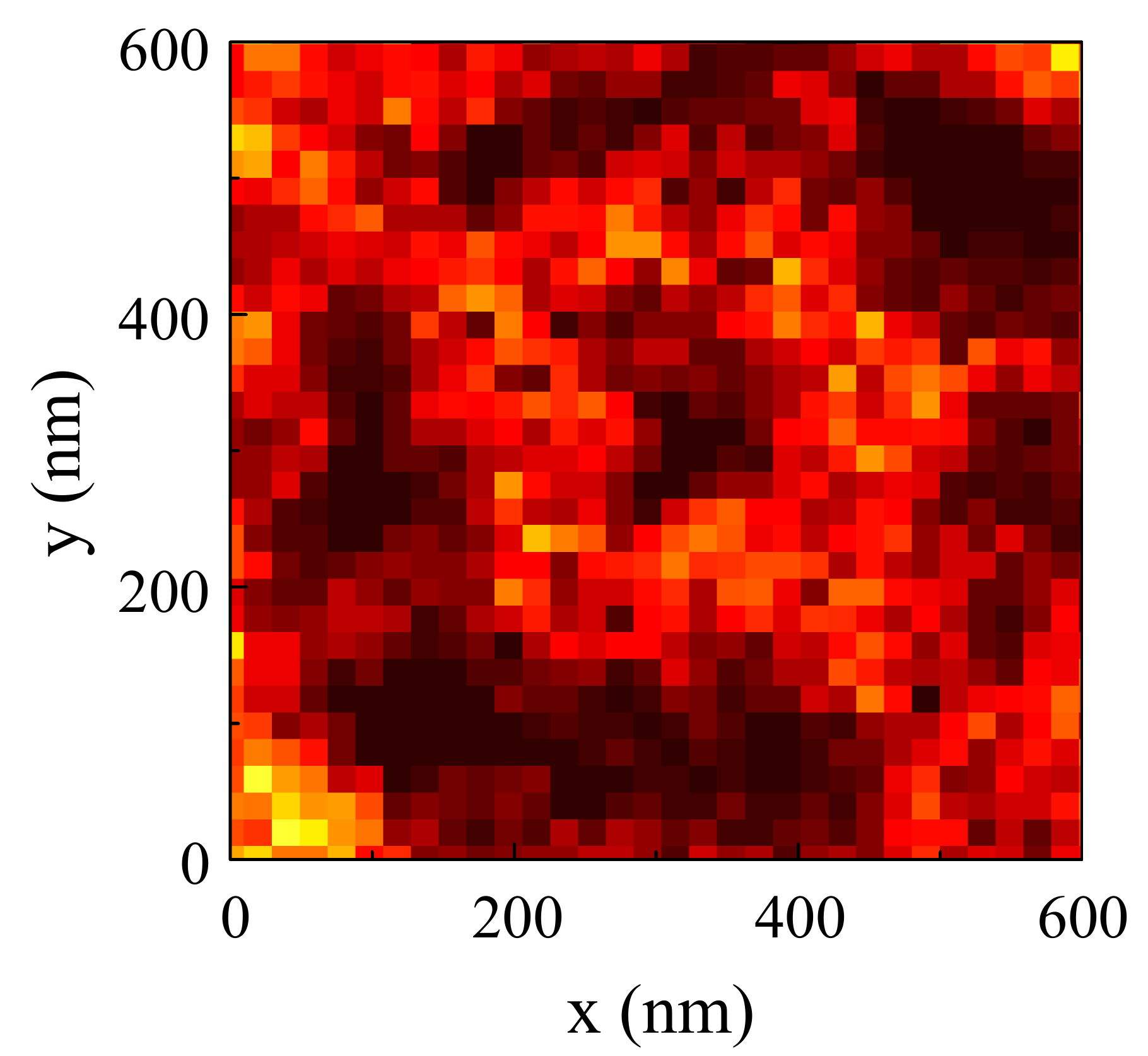}
		\caption{\label{fig9d}}
		\end{subfigure}
\begin{subfigure}{5.0cm}
		\includegraphics[width=0.99 \textwidth]{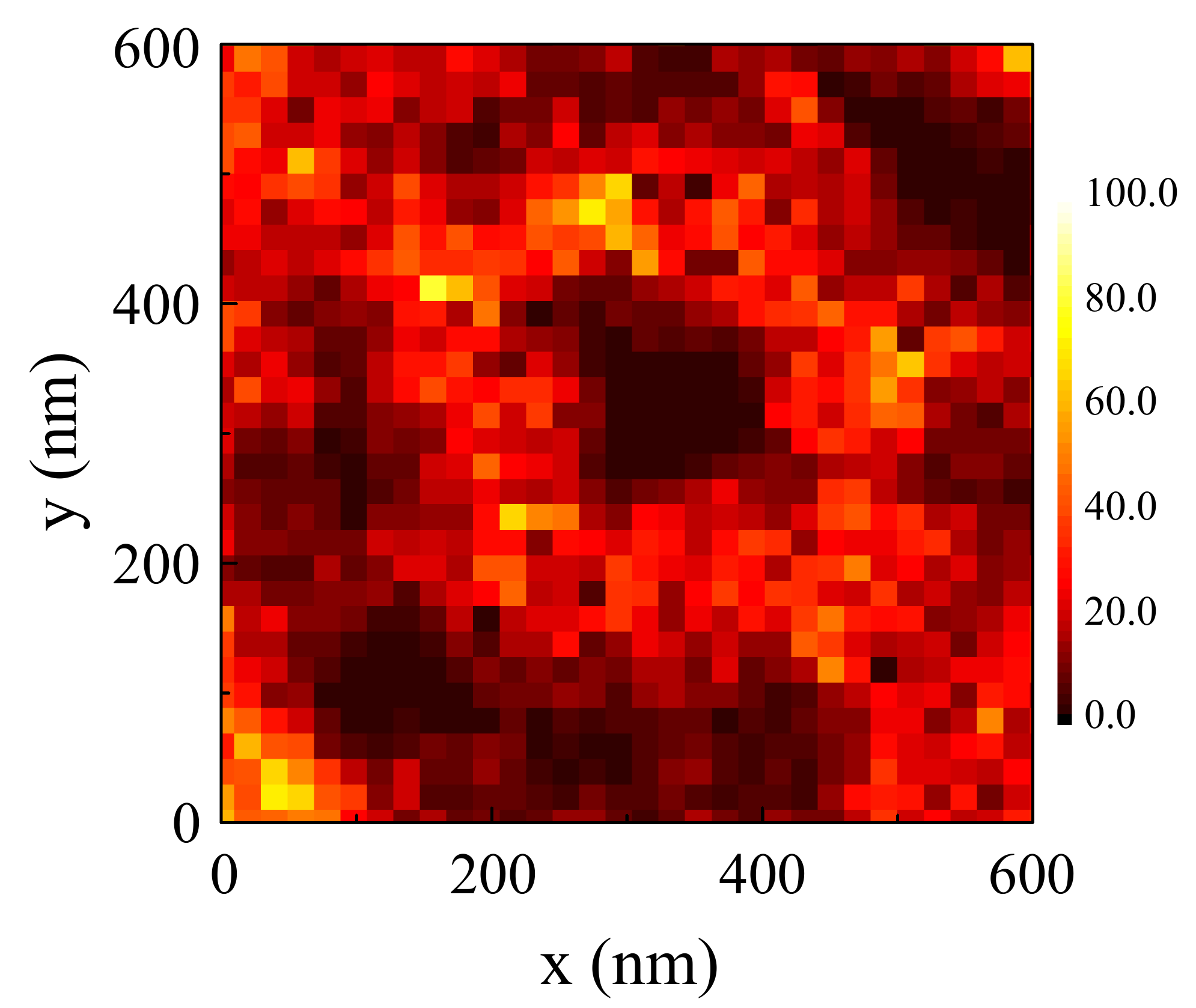}
		\caption{\label{fig9e}}
		\end{subfigure}
\begin{subfigure}{6cm}
		\includegraphics[width=0.99 \textwidth]{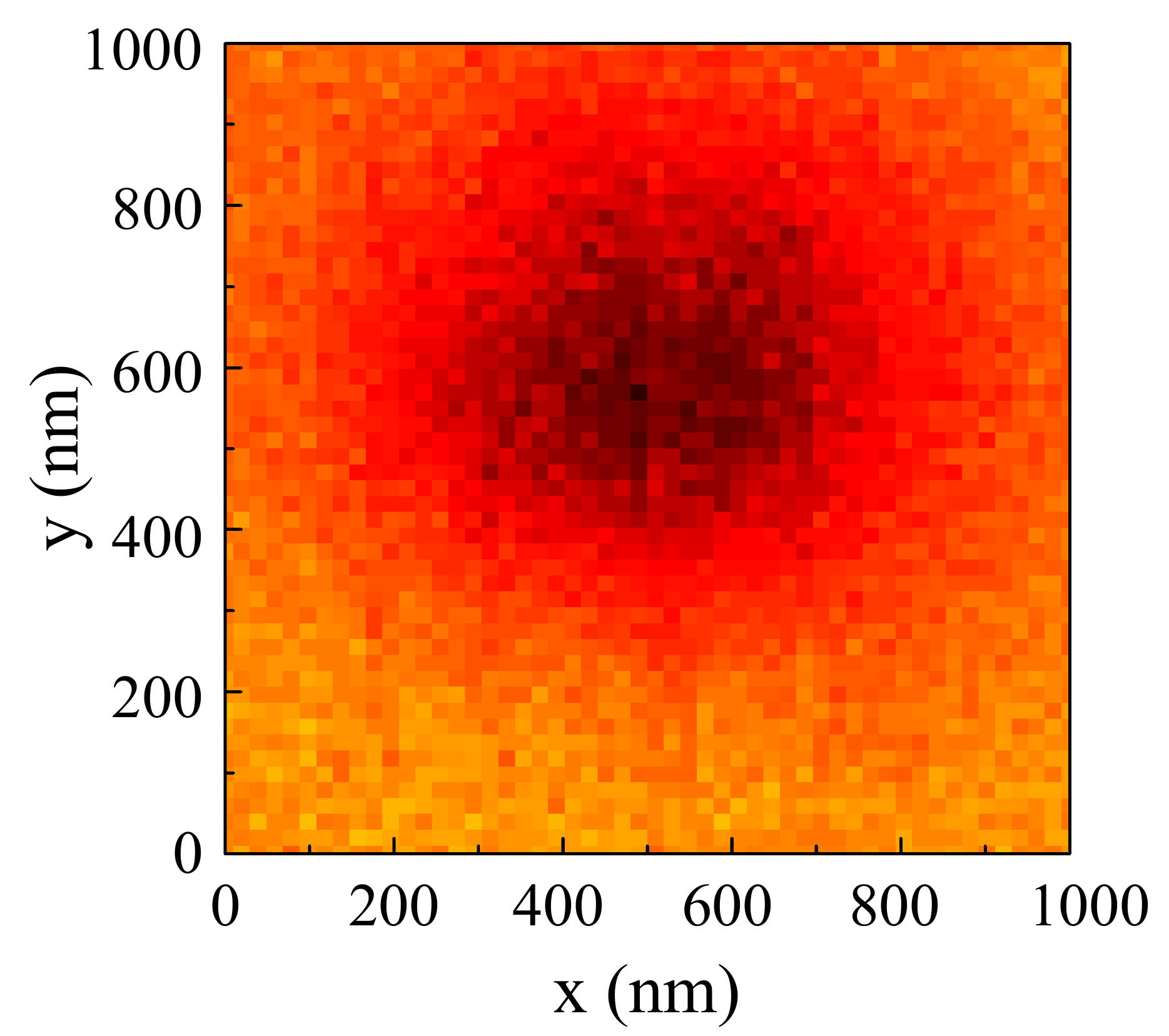}
		\caption{\label{fig9f}}
		\end{subfigure}
		\begin{subfigure}{6.5cm}
		\includegraphics[width=0.99 \textwidth]{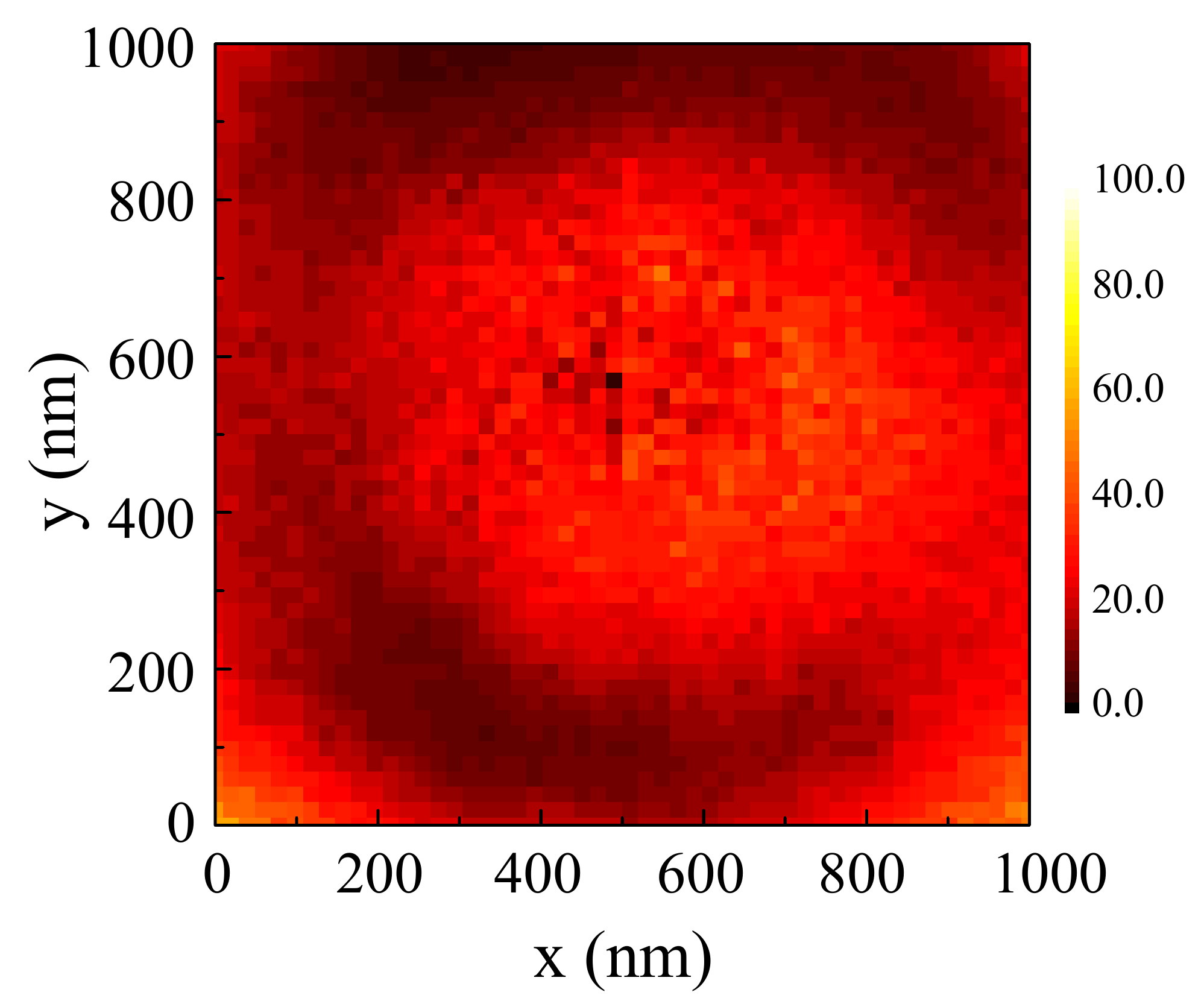}
		\caption{\label{fig9g}}
		\end{subfigure}
		\caption{ \textbf{Deconvolution of NSOM images:} The figure shows the results of deconvolution using the influence of tip as PSF for the NSOM scans. (a) shows the simulated field intensity profile for an isolated nanodisc for reference. (b) shows a raw NSOM scan of a gold nanodisc of 150 nm diameter and 35 nm height. (c-e) show the deconvolution results using different scaling of PSF - 0.25 (c), 0.275 (d) and 0.3 (e). (f) displays a raw NSOM scan of a gold nanodisc of 250 nm diameter and 50 nm height. (g) is produced by deconvolution of (f) using a rescaled tip PSF of 0.98. The scale for (b-e) is given in (e) and for (f-g) is given in (g).}
		\label{fig9}
\end{figure*}

	In order to understand the effect of the gold coated fiber tip on the near field generated by the nanodisc and thus to deconvolute the measured NSOM image to extract the actual near-field profile, we performed FDTD numerical simulations with and without the tip. For this, we included a model of the tip in the FDTD simulation setup (fig. \ref{fig2}). The tip was approximated as a frustum of a cone at 10 degrees from the vertical, with a Silicon Dioxide (SiO$_2$) core of 150 nm at the apex, and a uniform outer Au coating of 100 nm. The tip was positioned at a height of 5 nm above the nanodisc, and its position varied in an area of 200 nm$\times$200 nm. The entire region was simulated for each position of the tip as if the tip was scanning over the surface of the nanodisc. Two monitors were placed to record the electric field intensity - one on the surface of the nanodisc directly below the tip, and the other at the center of the NSOM tip. Fig. \ref{fig8} shows the electric field intensity as expected in the NSOM scan. A comparison of the without (Fig. \ref{fig9a}) and with tip results shows that the presence of the tip causes a distortion in the near-field of the nanodisc.\par

	To extract the actual near-field profile from the NSOM measurements, we need to deconvolute the measured profiles with the Point Spread Function (PSF) corresponding to the tip. We first attempted to remove the influence of the tip from the NSOM scan image via Blind Deconvolution in MATLAB. We presume that the tip's electric field response causes distortion in the actual response of the nanodisc (Fig.\ref{fig8}). Hence, the intensity profile captured by the NSOM (Fig.\ref{fig6}) can be modelled as a convolution of a function that represents the distortion caused by the tip and an isolated nanodisc. Mathematically,

\begin{align*}
  i_{NSOM} &= \sum_{all\ pixels} I_{tip}*I_{disc}
\end{align*}
where $i_{NSOM}$ denotes the NSOM scan, and $I_{tip}$ and $I_{disc}$ denote the Fourier transforms of the intensity profile of distortion from tip and nanodisc sample, respectively. Hence, to remove the effects of the NSOM tip on the near-field of the nanodisc sample, we deconvolve the distortion profile from the NSOM scan, i.e.

\begin{align*}
  i_{disc} &= \sum_{all\ pixels} \frac{I_{NSOM}}{I_{tip}}
\end{align*}
where $i_{disc}$ denotes the intensity profile of the nanodisc sample, and  $I_{NSOM}$ and $I_{tip}$ denote the Fourier transforms of the NSOM scan image and distortion from the tip, respectively. A blind deconvolution did not result in good match between the simulated and deconvoluted NSOM image.

 In order to obtain the distortion profile due to the tip accurately, we deconvoluted the simulated intensity profile of an isolated nanodisc from the simulated intensity profile with tip placed over a nanodisc. The result of this deconvolution gives the distortion of field intensity around the nanodisc due to the presence of the tip. But this is not enough to capture the effect of the tip scanning over the nanodisc. In order to accommodate the effects of different positions of the tip around the disc, the distortion profile is rotated by 90$^{\circ}$, 180$^{\circ}$ and 270$^{\circ}$, and all three profiles are added to the original one to obtain the PSF used for the deconvolution.  Following this, iterative Blind Deconvolution algorithm is applied to the NSOM scan image using $i_{tip}$ as the PSF, and $i_{NSOM}$ as the `blurred' image function to obtain deconvoluted image ($i_{disc}$) of the near field of the sample. Fig. \ref{fig9} shows the NSOM scan image before and after the deconvolution, as well as a simulated field profile for comparison. The deconvolved NSOM scan image shows that the central maxima deconvolve into a ring-like structure as expected from simulated results.\par

	The deconvolved image depends on the PSF and number of iterations used in the algorithm. We found that the deconvolved image is closest to simulated results when we scale the PSF down to 0.275 in comparison to its initial size. This scaling can be explained by considering that the tip is bent and is non-uniform, in terms of shape and gold-coating. Fig. 8 shows the dependence of the deconvolved image on the scaling of the PSF from 0.25 - 0.3, which gives the optimal results. This PSF, depending on its scaling, can be used to deconvolute NSOM images taken with tips of different sizes. Fig. 8(f-g) shows the NSOM scan image and deconvoluted image of the same nanodisc taken with a 250 nm tip. This clearly shows that the same PSF with different scaling can be used to deconvolve images to display the field intensity distribution around an isolated nanodisc, within the limits of the resolution of each individual tip. Since the resolution itself depends on the size of the aperture of the tip, the deblurring of images would be less effective as the size of the aperture increases.
	
	In conclusion, we have shown the generation of optical near field potential due to nanodisc for the trapping of a cloud of ultracold atoms. We also study the feasibility of generation of periodic potentials using this kind of hybrid system for a broad range of atomic species. With the ability to enhance the nearest neighbor tunneling rate, this kind of system can open up new domains of many exotic phases of matter. Finally, we have also done the qualitative study of the effect of gold coated nanotip on the field profile generated by the nano pillars. Apart from this, our deconvolution strategy may be adopted in order to deconvolute complex near-field maps in order to remove the influence of the tip on the near-field of the isolated system. This kind of an approach would depend on the shape and size of probes used, apart from the probing method. But this procedure cleans up NSOM images and helps highlight the relevant features hidden in such scans by deblurring. \par
	
	\begin{acknowledgement}
	The authors acknowledge funding support by the Department of Science and Technology (DST, Govt. of India) for grants through EMR/2014/000365 and the DST Nanomission Thematic Unit Program. SK would like to thank the Council for Scientific and Industrial Research, Govt. of India for research fellowship.
	
		
		
	\end{acknowledgement}
	

\providecommand{\latin}[1]{#1}
\makeatletter
\providecommand{\doi}
  {\begingroup\let\do\@makeother\dospecials
  \catcode`\{=1 \catcode`\}=2 \doi@aux}
\providecommand{\doi@aux}[1]{\endgroup\texttt{#1}}
\makeatother
\providecommand*\mcitethebibliography{\thebibliography}
\csname @ifundefined\endcsname{endmcitethebibliography}
  {\let\endmcitethebibliography\endthebibliography}{}

\end{document}